\documentclass[12pt,preprint]{aastex}
\usepackage{amsmath}

\newcommand \lya{Ly$\alpha$ }
\def\dim#1{\mbox{\,#1}}

\begin{document}

  \title{AGN outflows in a Cosmological Context}
  
  \author{Robyn Levine and Nickolay Y. Gnedin}
  \affil{Center for Astrophysics \& Space Astronomy, University of Colorado, Boulder, CO 80309}
  \email{robyn.levine@colorado.edu, gnedin@casa.colorado.edu}

  
  \begin{abstract}
    
  We combine cosmological simulations with an AGN luminosity function
  constrained by optical surveys in order to create a realistic AGN
  distribution in which to model AGN outflows.  The outflows are modeled
  with assumptions of spherical symmetry and energy conservation.  We study
  the importance of the kinetic luminosity of AGN outflows in determining
  the fractional volume of the IGM filled with outflows as a function of
  redshift.  We find that kinetic luminosities of $< 10\%$ of the
  bolometric luminosities are required or the entire IGM would be filled
  with outflows at $z=0$.  We also examine the effects of varying the AGN
  lifetime and the bias parameter, which describes how tightly correlated
  AGN are to high density regions.  We find that longer lifetimes
  ($\tau_{AGN} \sim 1\dim{Gyr}$), as well as small bias parameters increase
  the filling fraction.

  \end{abstract}

  \keywords{cosmology: theory---galaxies: active---intergalactic
  medium---large-scale structure of universe---quasars: general}


  \section{INTRODUCTION}

  Outflows from active galactic nuclei (AGN) potentially play a very
  significant role in the evolution of large-scale structure.  Such
  outflows consist of hot, tenuous out-flowing gas detected in absorption,
  and the powerful radio jets detected in some quasars.  Blue-shifted
  absorption lines (relative to systematic velocity of the AGN), such as
  \lya and the C {\sc iv} and N {\sc v} doublets, indicate gas moving away
  from the central source at high velocities (Ulrich 1988; Crenshaw et
  al. 1999).  Broad absorption lines (BAL) indicating even higher
  velocities, are detected in luminous AGN (e.g., Hewett \& Foltz 2003).
  Radio-loud quasars (RLQ) also carry substantial amounts of energy into
  the intergalactic medium (IGM) via collimated jets of relativistic plasma
  (e.g., Begelman, Blandford, \& Rees 1984).  AGN outflows may play a role
  in distributing magnetic fields into the IGM (e.g., Furlanetto \& Loeb
  2001), and they can impact the evolution of their host galaxies by, for
  example, regulating the growth of supermassive black holes (Wyithe \&
  Loeb 2003).  BAL outflows and RLQ are also possible mechanisms for
  heating the intra-cluster medium (e.g., Valageas \& Silk 1999; Nath \&
  Roychowdhury 2002).  The magnitude of the influence of outflows on the
  IGM depends on a few key properties of AGN, such as the relationship
  between kinetic and bolometric luminosity, that still need to be
  constrained by observation.

  In RLQ, the kinetic luminosity of the jet is thought to be correlated
  with the bolometric luminosity (Willott et al. 1999).  Direct estimates
  of the kinetic luminosity in BAL outflows can be made if such quantities
  as the covering fraction of the outflows, the outflow velocity, the
  radius of the outflow, and the column density are known.  The velocities
  are accurately obtained from the absorption lines, and the covering
  fractions can be estimated with statistical arguments (Weymann 1997).
  The radius of the outflow can be obtained through observations of the
  photoionizing flux of the central source (Krolik 1999).  Estimates of the
  column densities can be made by studying the UV and X-ray absorption
  lines (Gallagher et al. 1999, 2001).  Uncertainties in the above
  quantities translate to uncertainties in the kinetic luminosity, and
  therefore into uncertainties in the energy of AGN outflows.  More
  energetic outflows can fill a larger volume, having a greater impact on
  large-scale structure.
  
  In this paper, we model the growth of AGN outflows using energy
  conservation arguments and simple approximations about their geometry.
  Through numerical simulations we model the distribution of these outflows
  and hence we can estimate the degree to which they affect the universe on
  global scales, or the fractional volume occupied by AGN outflows.  In \S
  \ref{sec:sim}, we describe some of the details of the cosmological
  simulation and the luminosity function that we have combined to obtain an
  AGN distribution.  In \S \ref{sec:out} we explain the assumptions
  surrounding our physical model of the growth of heated bubbles around AGN
  as well as our selection of kinetic luminosity.  In \S \ref{sec:ana}, we
  show the advantage of using a simulated density distribution over using
  simple statistical methods for the distribution of AGN.  In \S
  \ref{sec:res} we examine the effects of simulation size and resolution,
  AGN lifetime, and AGN bias on the cumulative effect of outflows, and we
  conclude in \S \ref{sec:con}.  For the rest of this paper, we assume
  $\Omega_m = .27$, $\Omega_{\Lambda} = 0.73$, and $\Omega_b = .04$ with
  $\Omega_b h^2 = 0.02$, consistent with the WMAP data.
  
  
  \section{SIMULATING THE AGN ENVIRONMENT AND DISTRIBUTION}
  \label{sec:sim}
  
  In order to understand the influence of AGN outflows on a global scale,
  it it useful to model the outflows in the context of large-scale
  structures.  We assume here that AGN trace high density regions, and so
  use a gas density distribution to bias AGN in our study.  Other studies
  have, for example, approximated the distribution of AGN with statistical
  formulae.  Tegmark, Silk, \& Evrard (1993) and Furlanetto \& Loeb (2001;
  hereafter, F \& L) both use Poisson distributions to model the spatial
  distribution of galaxies for simplicity, in order to obtain filling
  fractions of supernova-driven winds and of AGN outflows respectively.  We
  combine a z-dependent luminosity function with a simulated gas density
  distribution, allowing us to consider the AGN in their appropriate
  environments rather than homogeneously distributing them throughout the
  universe.  In \S \ref{sec:ana}, we compare a Poisson distribution of AGN
  outflows with our model.


  \subsection{Cosmological Density Distribution}
  \label{subsection:distr}
  In order to model the evolution of the gas density distribution in the
  universe, we use a standard Particle-Mesh code to simulate the
  distribution of the dark matter, and we assume that on the scales we are
  considering, the gas distribution follows that of the dark matter, which
  is supported by cosmological gas dynamics simulations (Gnedin 2000; Chiu
  \& Ostriker 2000; Miller \& Ostriker 2001; Somerville 2002; Tassis et
  al. 2003; Benson \& Madau 2003; Susa \& Umemura 2004; Shapiro, Iliev, \&
  Raga 2004; Mo \& Mao 2004). We also assume that the temperature of the
  cosmic gas is constant at 15,000 K. This assumption is, clearly, an
  oversimplification, since the temperature of the cosmic gas is known to
  evolve with time (Ricotti, Gnedin, \& Shull 2000; Schaye et al. 2000;
  McDonald et al. 2001; Kim, Cristiani, \& D'Odorico 2002; Theuns et
  al. 2002; Hui \& Haiman 2003) and vary in space. However, since the
  typical sizes of AGN-driven bubbles are significantly larger than the
  scales over which the temperature of cosmic gas changes, this
  approximation is sufficient for our purposes.


  \subsection{Luminosity Function}
  \label{sec:lum}
    
  The above simulation produces a gas density distribution at each time
  step, from $z\sim 19$ to $z=0$.  In each step, we use a luminosity
  function to populate the simulation with AGN over a range of
  luminosities.  In order to place AGN into our simulation, we must
  implement an AGN luminosity function that applies to high redshifts.  The
  scarcity of high-z galaxy detections in surveys makes luminosity function
  predictions difficult, although the ability to make detections is
  improving.  We use a model by Schirber \& Bullock (2003) for the QSO
  luminosity function.  The model satisfies all existing constraints from
  optical surveys, such as the Sloan Digital Sky Survey (SDSS) data and the
  Great Observatories Origins Deep Survey (GOODS) data for $z>3$ (Fan et
  al. 2001a, 2001b; Cristiani et al. 2004) and Two Degree Field (2dF) data
  for $z< 2.3$ (Boyle et al. 2000).  It should be noted that optical
  surveys perhaps underestimate the faint-end, low-z luminosity function.
  Hard X-ray detections of AGN yield a higher number of faint AGN at
  low-redshift than in previous optical surveys (Barger et al. 2005).  The
  model that we use for now parameterizes the following luminosity
  function:
  
  \begin{equation} \label{eq:lum}
    \phi(L_B,z) = \frac{\phi_* / L_*}{(L_B/L_*)^{\gamma_f}+(L_B/L_*)^{\gamma_b}}\;.
  \end{equation}

\begin{figure}[hpt] 
    \centering
    \epsscale{.7}
    \plotone{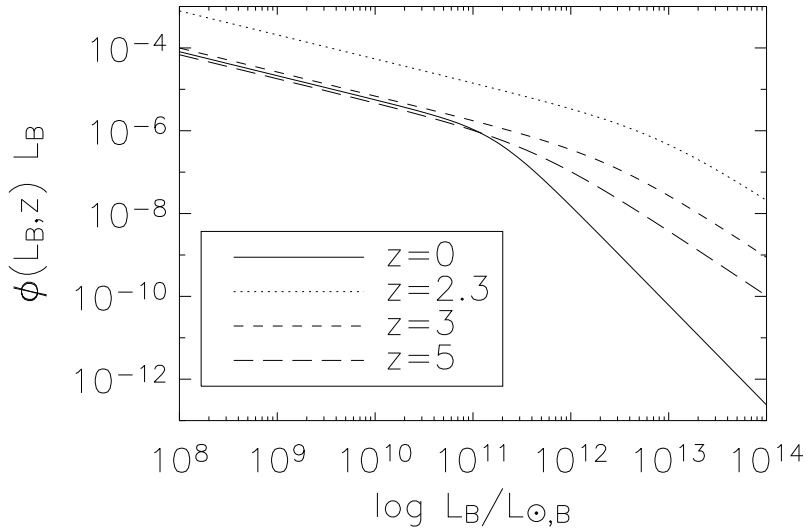}
    \caption{{\label{fig:lumf}} Calculated comoving number density of AGN
      at a given redshift, using a luminosity function consistent with 2dF,
      SDSS and GOODS data.}
  \end{figure}

  \noindent In the above equation, $\phi_*$ is the average comoving number
  density of AGN, $L_*$ and $L_B$ are the break luminosity and AGN B-band
  luminosities respectively (given in units of $L_{\sun,B}$ where we have
  followed S \& B in using $2.11 \times 10^{33}\dim{ergs s}^{-1}$ for the
  B-band luminosity of the sun), and $\gamma_f$ and $\gamma_b$ are the
  faint and bright-end slopes respectively.  As in S \& B, we interpolate
  the SDSS and 2dF fits for $2.3<z<3$.  The parameterization is given in
  Table \ref{tb:par}, and the luminosity function for several different
  redshifts is plotted in Figure \ref{fig:lumf}.  Note that for $z>3$,
  $L_*$ and $\phi_*$ depend on the weighted emissivity of AGN (accounting
  for contributions to the ionizing rate from other sources, such as
  stars).  The weighted emissivity is given by $\log_{10}\
  \hat{\varepsilon}^Q = -0.245 z + 0.596$, a simple power law fit to the
  values given in S \& B Table 1.  We have chosen a minimum AGN luminosity
  of $10^8 L_{\sun,B}$ as in S \& B, who argue that this represents the
  faintest Seyfert galaxies found.  We extrapolate the high-z
  parameterization of Table \ref{tb:par} to $z \sim 19$, which introduces a
  significant uncertainty in the abundance of AGN at high redshift.
  However, as we show below, AGN at $z<3$ dominate, so this uncertainty
  affects our results insignificantly.  The exact redshift range for each
  run depends on the size of the simulation box since larger boxes can
  sample an earlier, rarer population of AGN.

  \begin{deluxetable}{cccc}
    \tablecolumns{5} \tablewidth{0pt} \tablehead{\colhead{} &
    \colhead{$z<2.3$} & \colhead{$2.3<z<3$} & \colhead{$z>3$} }
    \tablecaption{{\label{tb:par}} Parameterization of luminosity function}
    \startdata $\log_{10}$ ($\phi_{\star}\dim{Gpc}^3$) & $3.029$ & $-0.87 z +
    5.02$ & $2.80 + 2.72\ \log_{10}\ \hat{\varepsilon}^Q(z) + 0.81\ (z-3)$
    \\ $\log_{10}\ (L_{\star}\ L_{\sun,B}^{-1})$ & $11.24 + 1.36 z - 0.27
    z^2$ & $-0.72 z + 14.6$ & $12.2 - 1.72\ \log_{10}\
    \hat{\varepsilon}^Q(z) - 0.81\ (z-3)$ \\ $\gamma_b$ & $3.41$ & $-1.19\
    z+6.14$ & $2.58$ \\ $\gamma_f$ & $1.58$ & $1.58$ & $1.58$ \\ \enddata
  \end{deluxetable}

  It should be noted that we used S \& B's ``Model A'' to determine this
  particular luminosity function for $z \geq 3$. The model uses the
  emissivity of the ionizing background, assuming a stellar as well as an
  AGN contribution, to constrain the AGN luminosity function in combination
  with the SDSS data.  The model implements the ionizing rates of McDonald
  \& Miralda-Escud\'{e} (2001) and assumes an escape fraction of $0.16$
  (including the stellar contribution to the ionizing background).  The
  model neglects optically obscured AGN on the basis that they do not make
  a significant contribution to the ionizing background.  This means that
  the luminosity function we use could be excluding some AGN that host
  outflows.  Of the models described by S \& B, we have chosen `Model A''
  because it is most consistent (of the models studied by S \& B) with the
  faint end luminosity function (for $z>4$), from GOODS (Cristiani et
  al. 2004).

  We use the luminosity function to calculate the number of AGN at a given
  redshift and luminosity in each simulation box, and we then place each
  one into a random location, but with a bias toward high density regions.
  In order to distribute AGN so that they correspond to regions of high
  density, we calculate, for each cell, the probability of hosting an AGN
  given by
  
  \begin{equation} \label{eq:prob}
    P(i,j,k) = \frac{\rho_m^{\alpha}(i,j,k) \Delta V}{\sum\limits_{i,j,k = 0}^{N-1} (\rho_m^{\alpha} \Delta V)}\;.
  \end{equation}

  \noindent Above, $\rho_m$ is the matter density (in units of average
  baryon density) in each cell specified by coordinates $i$,$j$,$k$; $N$ is
  the size of simulation box; $\Delta V$ is the comoving volume of each
  cell; and $\alpha$ is a bias parameter ensuring that regions of higher
  density are more likely to host AGN.  The above probability function is
  independent of the characteristics of individual AGN (such as
  luminosity).  Analysis of existing observations suggests that AGN are
  more biased toward high density regions at increasing redshifts (e.g.,
  Croom et al. 2004; Porciani, Magliocchetti, \& Norberg 2004).  In \S
  \ref{subsection:bias}, we will examine the effects of different bias
  parameters on our results.  In order to place the AGN in specific cells,
  we randomly choose a cell and compare the value of $P(i,j,k)$ in that
  cell to a randomly generated number between zero and one.  If $P(i,j,k)$
  is greater than the random number, an AGN goes into the cell.  Otherwise,
  the process repeats until the AGN has been placed in the simulation.
  This method has the effect of biasing AGN toward regions of higher
  density, while fully allowing for Poisson noise in their distribution.


  \section{AGN OUTFLOWS}
  \label{sec:out}

  The following sections detail the distribution and kinematics of AGN
  outflows in our model.  Outflows are not observed in all AGN, and so \S
  \ref{subsection:dist} deals with our method of selecting AGN to host
  outflows.  In \S \ref{subsection:evol}, we describe our assumptions about
  the expansion and evolution of individual outflows into the IGM, and in
  \S \ref{subsection:kinfrac}, we discuss the effects of kinetic luminosity
  on the outflows.
  
  \subsection{Distribution of Outflows}
  \label{subsection:dist}

  In our model we assume that the AGN produce outflows that are
  responsible for distributing tenuous, hot gas into the IGM.  The outflows
  may also be a mechanism for depositing metals and magnetic fields into
  the IGM.  However, outflows are only associated with a fraction of AGN,
  which we must reflect in our model.  Detection of blue-shifted broad
  absorption lines in an AGN's spectrum indicates out-flowing gas from the
  nucleus.  Until fairly recently, observations indicated that these broad
  absorption lines were limited to radio-quiet quasars (Stocke et al.
  1992), and only seen in $\sim 10\%$ of them.  However, recent detections
  of BAL outflows in RLQ (e.g., Brotherton et al. 1998; Menou et al. 2001)
  suggest that BAL outflows do not necessarily follow this radio dichotomy.
  There is, however, evidence for some luminosity dependence in the
  occurrence of BAL outflows.  Crenshaw et al. (1999) examined the UV
  spectra of several Type I Seyfert galaxies obtained with the Hubble Space
  Telescope, and found intrinsic, narrow absorption lines in more than half
  of their sample.  The absorption spectra of these low-luminosity objects
  show similarities to the BAL features in high-luminosity objects,
  suggesting a possible relationship between the two.  On the higher
  luminosity end, Hewett \& Foltz (2003) apply a magnitude correction to
  the Large Bright Quasar Survey (LBQS; Hewett, Foltz, \& Chaffee 1995,
  2001) and find a higher percentage of BAL quasars than previously
  determined.  Magnitude and flux-limited samples can exclude BAL quasars
  in which the absorbing gas reduces the spectral energy distribution of
  the quasars in the wavelength range of selection.  After applying a
  correction for this effect, Hewett \& Foltz find that $\sim 20\%$ of the
  AGN in the sample host outflows.

  In our model, we interpolate the above low and high-luminosity limits for
  the fraction of AGN hosting outflows, $f_{out}$, and combine this
  fraction with the luminosity function of \S \ref{sec:lum}.  Our
  formulation is as follows:

  \begin{equation} \label{eq:out}
    f_{out}=
    \begin{cases}
      .5 \ \ \ \ \ \ \ \ \ \ \ \ \ \ \ \ \ \ \ \ \text{for $\log L_B < 10$}\;, \\
      1.5 - 0.1\ \log L_B \ \ \text{for $10 \le \log L_B \le 13$}\;, \\
      .2 \ \ \ \ \ \ \ \ \ \ \ \ \ \ \ \ \ \ \ \ \text{for $\log L_B > 13$}\;,
    \end{cases}
  \end{equation}

  \noindent where the lower and higher-end luminosities are consistent with
  the fractions determined above.  It is not well known whether the above
  statistics also apply to the population of X-ray detected AGN without
  optical counterparts.  However, because X-ray surveys detect more
  low-luminosity AGN than optical surveys, it is possible that there are
  more AGN containing outflows than we here assume.  Additionally, it
  should be noted that because of the range of covering fractions of AGN
  outflows, the fraction of AGN hosting outflows could be larger than what
  is observed, even in optical surveys, due to orientation effects (Morris
  1988; Weymann et al. 1991; Hamann, Korista, \& Morris 1993).  Therefore,
  our assumptions should provide a conservative lower limit to the number
  of AGN hosting outflows.

  
  \subsection{Evolution of Outflows}
  \label{subsection:evol}
  
  In order to understand the degree to which active galaxies affect the
  IGM, it is important to have an accurate model of the expansion of hot,
  ionized gas into the IGM.  The environment of the outflow must be
  considered in order to model the expansion.  Using the density profile
  described in \S \ref{sec:sim}, we can, to a degree, reproduce the
  environment surrounding each of the individual active galaxies that we
  placed in our simulation in accordance with the luminosity function
  (eq. \ref{eq:lum}) and $f_{out}$ (eq. \ref{eq:out}).
  
  The lifetime of the active galactic nucleus is short compared to the
  expansion time of the bubbles, which lasts over the duration of our
  simulation.  Therefore, we consider the active phase as a brief energy
  injection, followed by an adiabatic blast wave gathering up material in
  the IGM, analogous to the adiabatic phase of a supernova remnant.  As in
  Scannapieco \& Oh (2004), we have used the familiar Sedov-Taylor blast
  wave model to determine the size of the bubbles in our analysis.  Their
  Equation 10, which we adopt for convenience of units, follows:
    
  \begin{equation} \label{eq:sed}
    R_s = (1.7\dim{Mpc}) \left(\frac{E_{k}}{10^{60}
    \dim{ergs}}\right)^{1/5} (1+\delta_m)^{-1/5} (1 + z)^{-3/5}
    \left(\frac{t_{age}}{10^9\dim{yr}}\right)^{2/5}\;.
  \end{equation}

  \noindent In the above equation, $E_k$ is the kinetic energy injected by
  the AGN, $\delta_m$ is the overdensity, and $t_{age}$ is the time since
  the active phase began.  The overdensity is an average over all cells
  within the radius of the bubble, obtained using an iterative technique.
  The above Sedov-Taylor solution, with a constant density, is clearly not
  valid as the outflow escapes its host galaxy and travels first through a
  region with a steeply-falling off density profile, e.g. the NFW profile
  (Navarro, Frenk, \& White 1997).  However, our simulation does not
  resolve this stage of the expansion.  The virial radius of a typical
  $L_\star$ galaxy is well below the resolution of our simulation.  Without
  modeling the expansion of the outflows inside this region, there is
  uncertainty in the speed with which the outflows escape their host
  galaxies.  However, this uncertainty is hidden by the parameter $E_k$,
  which describes the actual energy input from the AGN.

  We assume that the bubbles expand according to the above equation until
  they reach a pressure equilibrium with their environment.  If pressure
  equilibrium is reached before the energy injection has stopped, the
  growth of the bubbles is determined by the surrounding pressure and the
  injected energy, and the size is given by

  \begin{equation} \label{eq:pres}
    R_P = (3.24 \times 10^{-25}\dim{Mpc}) \left(\frac{3E_k}{4\pi P}\right)^{1/3}\;. 
  \end{equation}

  \noindent The pressure of the surrounding medium is given by $P =
  (1+\delta_m) \bar{n}_b k_b T$, where the quantity $(1+\delta_m)
  \bar{n}_b$ is the average gas density inside the bubble as determined
  above, and $T=1.5 \times 10^4\dim{K}$ (the average temperature of the
  IGM).  The kinetic energy in Equations \ref{eq:sed} and \ref{eq:pres} is
  given by $E_k = \varepsilon_{kB} L_B t_{age}$ where $t_{age}$ is the age
  of the AGN during the active phase, or the lifetime of the AGN once the
  active phase has ended.  We assume a constant lifetime of $10^8\dim{yrs}$
  for all AGN until \S \ref{subsection:life} in which we will examine other
  lifetimes. The parameter $\varepsilon_{kB}$ is given by the ratio of the
  kinetic luminosity of the outflow to the AGN B-band luminosity
  ($L_k/L_B$).  Each AGN injects kinetic energy into the IGM at a rate
  ($L_k$) correlated with the AGN's luminosity.  We assume that the AGN are
  accreting at roughly their Eddington rates, so that $L_{edd} \sim
  L_{bol}$.  As in F \& L, we adopt $L_{bol} \sim 10 L_B$, consistent with
  Elvis et al. (1994).  We describe our choice of the kinetic fraction,
  $L_k/L_{bol}$ or $\varepsilon_k$, in the next section.

  After the energy injection phase, bubbles in pressure equilibrium no
  longer expand as a result of the energy injection of the AGN.  Any
  subsequent evolution of the bubbles in our simulation is determined by
  the Hubble expansion and the evolution of the density distribution in the
  surrounding environment.  Overlap of the bubbles is not likely to affect
  the expansion significantly.  When an expanding bubble in pressure
  equilibrium overlaps with the interior of another bubble, encountering
  densities much lower than those typical in the IGM, the expansion speed
  does not change, as there is no longer a pressure gradient.

  Our model does not include radiative cooling, as the cooling times for
  these bubbles are typically much longer than the timescales we consider.
  However, for bubbles located in cluster environments (higher densities
  and temperatures, etc.), radiative cooling, as well as other physical
  processes, may become important.  Accurately modeling outflow physics in
  these complex environments requires the use of hydrodynamical
  simulations.  Separate work is currently underway to understand the
  impact of AGN outflows in these environments (e.g., Br\"{u}ggen \& Kaiser
  2002; Ruszkowski, Br\"{u}ggen, \& Begelman 2004).
  
\begin{figure}[hpt] 
    \centering
    \epsscale{1.0}
    \plotone{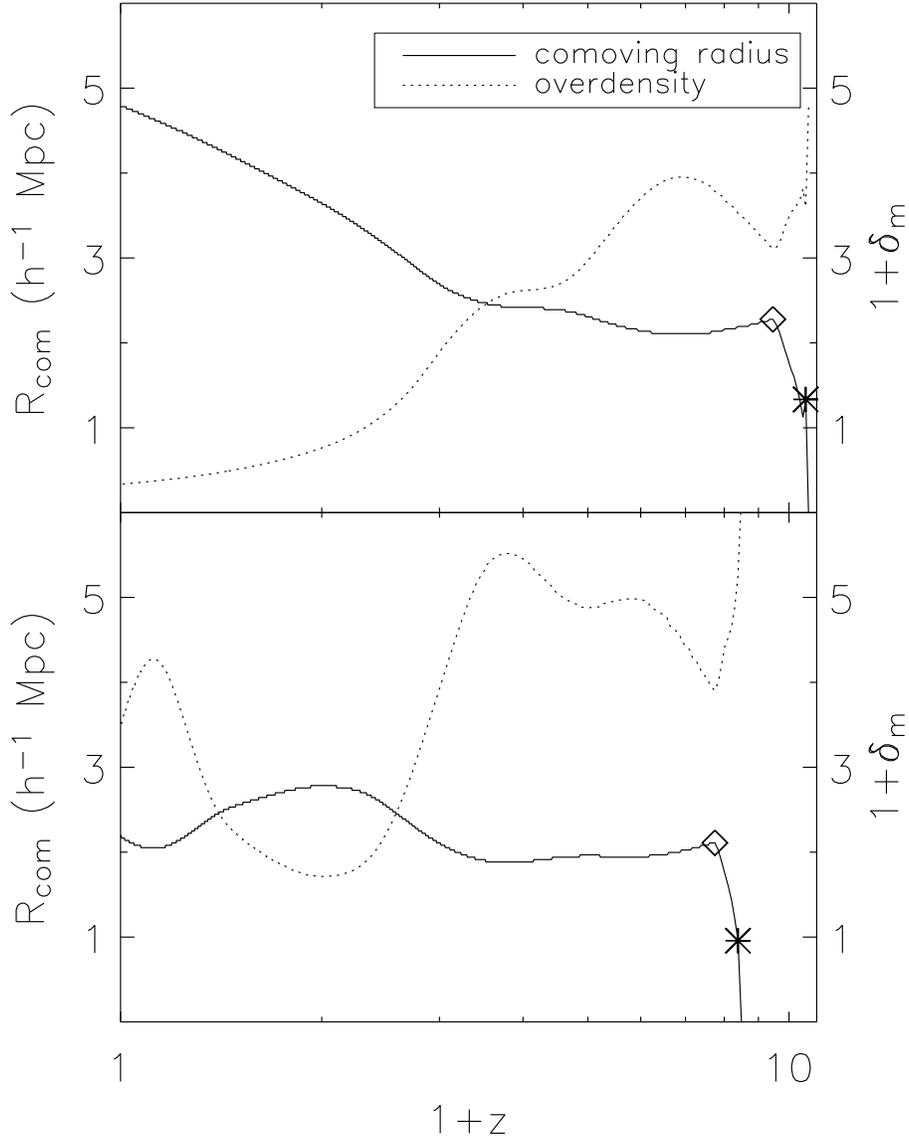}
    \caption{\label{fig:comr} Overlay of the comoving radius of two AGN
      bubbles ({\it solid}) in different environments, and the average
      overdensity, $1+\delta_m$, inside each bubble ({\it dotted}).  We
      have chosen AGN with luminosities of $\sim 10^9 L_{\sun,B}$.  We show
      the point at which the bubbles reach pressure equilibrium with their
      environments ({\it star}) and the end of the active phase after a
      lifetime of $10^8\dim{yrs}$ ({\it diamond}).  The figures represent
      typical AGN, although the exact behavior is a function of luminosity
      and environment.  Note that the continued growth of the bubble after
      reaching pressure equilibrium is a result of the continuing energy
      injection from the nucleus.  Once the energy injection (active phase)
      ends, the bubble size only changes with the evolution of its
      environment and the Hubble flow.  In the top figure, the bubble is
      expanding into a region of lower density, allowing the expansion to
      continue.  In the bottom figure, the bubble expands into a denser
      region, eventually halting the expansion.}
  \end{figure}

  Figure \ref{fig:comr} shows the evolution of the bubble size for two
  typical AGN (each with luminosities of $\sim 10^9 L_{\sun,B}$) residing
  in different environments within the simulation.  The figure shows that
  the AGN residing in the lower density environment produces a larger
  bubble ($\sim 4.8\ h^{-1}\dim{Mpc}$) than the AGN in the higher density
  environment ($\sim 2.2\ h^{-1}\dim{Mpc}$).

  In both F \& L and Nath and Roychowdhury (2002), AGN outflows are treated
  as collimated jets that spread out into a cocoon after reaching pressure
  equilibrium at the end of the AGN's active phase.  Furthermore, both BAL
  AGN and RLQ are treated similarly, justified by the small covering
  fraction of BAL outflows, averaging at around $10\%$ (Weymann 1997).  We
  likewise adopt the practice of treating the two different objects
  similarly, noting that there is additional incentive in the likelihood of
  overlap between RLQ and BAL AGN.  Furthermore, because the time scales we
  consider are significantly longer than the AGN injection phase, we do not
  model collimated jets, but rather approximate the outflows as bubbles
  expanding into the IGM with spherical symmetry.  However, as demonstrated
  by Figure 2 of F \& L, if the energy injection is modeled assuming
  spherical symmetry during the active phase, rather than with collimated
  jets, the result is only a slightly smaller final comoving bubble size.
  The entirely spherical case produces a smaller cocoon because the surface
  area of the bubble causes it to decelerate sooner, whereas a collimated
  jet makes its way through the IGM more easily.

 
  \subsection{Kinetic Luminosity and Filling Fraction}
  \label{subsection:kinfrac}

\begin{figure}[hpt]
    \centering
    \epsscale{.7}
    \plotone{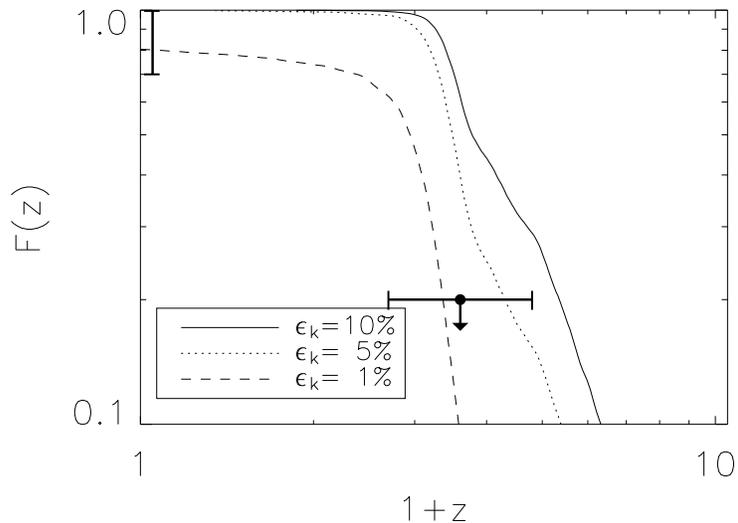}
    \caption{\label{fig:eps} Fractional volume filled with AGN outflows as
    a function of redshift for different kinetic fractions.  The bars
    represent the observational constraints on the filling fraction of AGN
    outflows from the \lya forest, as discussed in the text.}
  \end{figure}

  Estimates of the kinetic fraction, $\varepsilon_k$, are subject to a
  number of observational uncertainties surrounding BAL outflows.
  Observational constraints depend on quantities that are difficult to
  determine, such as the distance of outflows from the central source, and
  the covering fraction of the outflows (e.g., De Kool et al. 2001).  We
  use our model of AGN outflows to calculate the filling fraction of
  outflows as a function of redshift, $F(z)$, for a box of length $128\
  h^{-1}\dim{Mpc}$ (with $0.5\ h^{-1}\dim{Mpc}$ cells), assuming that all
  of the AGN are active for $10^8 \dim{yrs}$, and that they follow a
  constant bias, $\alpha=2$.  We then treat the kinetic fraction as a free
  parameter of our model.  We start with $\varepsilon_{kB} = 1$, or a
  kinetic fraction $\varepsilon_k=0.1$, as in F \& L.  Nath \& Roychowdhury
  (2002) argue that $\varepsilon_k=0.1$ is probably an upper limit for BAL
  outflows, assuming that the covering fraction of BAL outflows is $\sim
  10\%$.  Figure \ref{fig:eps} shows the fractional volume filled with AGN
  outflows for decreasing kinetic fraction.  We find that using a kinetic
  fraction of $10\%$, the entire simulation box is filled with outflows by
  $z\sim2$.  Similarly, Scannepieco \& Oh (2004) find that a kinetic
  fraction of $10\%$ overestimates feedback effects in their model of AGN
  outflows.

  Knowledge of the filling fraction of the \lya forest at various redshifts
  can provide further constraints, if we assume that the AGN outflows
  consist of hot, tenuous gas that cannot occupy the same volume of space
  as \lya absorbing regions.  At low-z, simple conclusions drawn from
  observations of the \lya forest constrain the fractional volume of voids
  to between $70$ and $99.6\%$ of the total volume, providing an upper
  limit to the filling fraction of AGN outflows (e.g., Penton, Stocke, \&
  Shull 2004; Dav\'{e} et al. 1999).  The vertical bar near $z=0$ in Figure
  \ref{fig:eps} shows the constrained filling fraction of voids.  An
  $\varepsilon_k$ of less than $10\%$ produces filling factors that are
  consistent with the above values, but imposing more precise constraints
  from \lya forest observations is difficult because of the range of column
  densities under consideration.  For somewhat higher redshifts
  ($1.7<z<3.8$), Duncan, Ostriker, \& Bajtlik (1989) have studied voids in
  the \lya forest and determined that voids with sizes between $10$ and
  $70\ h^{-1}\dim{Mpc}$ occupy $< 20\%$ of the volume.  The horizontal bar
  in Figure \ref{fig:eps} shows this upper limit for the filling fraction
  of AGN outflows over the appropriate redshift range.  We choose
  $\varepsilon_{kB} = .1$ ($\varepsilon_{k} = 1\%$) as our fiducial value,
  as it seems to match observational and theoretical constraints more
  closely than the higher values.


  \section{COMPARISON WITH A POISSON DISTRIBUTION}
  \label{sec:ana}

  As previously mentioned, it is possible to calculate the filling fraction
  of AGN outflows analytically by assuming that AGN are distributed
  according to Poisson statistics.  In this section, we demonstrate the
  advantage of including a realistically evolving density distribution in
  our model.

  We calculate the filling fraction of outflows using a Poisson
  distribution of sources and compare with our method of biasing AGN toward
  the high density regions within a cosmological simulation (see \S
  \ref{sec:sim}).  We first calculate the porosity of AGN outflows at each
  redshift, given by:

  \begin{equation}
    Q(z) = \frac{4\pi}{3} \int_z^{\infty}
    \frac{dz'}{\tau_{AGN}}\,\frac{dt'}{dz'} \int_{L_{min}}^{L_{max}} R^3
    \phi(L_B,z)\,f_{out}\,dL_B \;,
  \end{equation}

  \noindent where the above quantities are calculated in physical units.
  We then calculate the filling fraction assuming a Poisson distribution:

  \begin{equation}
    f(z) = 1 - e^{-Q(z)} \;.
  \end{equation}

\begin{figure}[hpt]
    \centering
    \epsscale{.7}
    \plotone{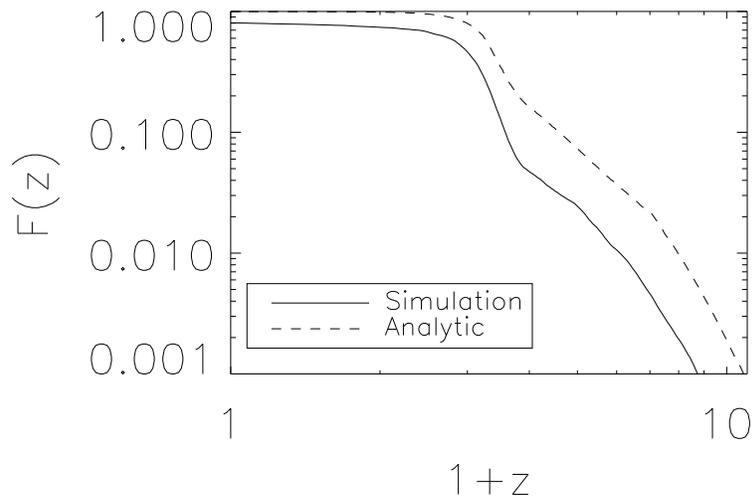}
    \caption{\label{fig:comp} Filling fractions for a uniform density
      distribution (no density bias) and for an evolving density
      distribution (with density bias).  The simple Poisson distribution
      results in a higher filling fraction than in the scenario accounting
      for an evolving density distribution.}
  \end{figure}

  \noindent In the above calculations, we have determined the volume of
  each outflow under pressure equilibrium conditions, and under the
  assumption that the energy injection is instantaneous.  The pressure is
  determined by the average particle density at each redshift and a
  constant temperature of $T = 1.5\times10^4\dim{K}$.  We compare the above
  filling fraction with that produced by our model, as described in the
  previous section, for a simulated volume of $128^3\ h^{-3}\dim{Mpc}^3$
  (with $0.5\ h^{-1}\dim{Mpc}$ cell resolution) in Figure \ref{fig:comp}.
  The Figure shows that the Poisson distribution of sources produces a
  higher filling fraction than our model.  The simulation provides a
  realistic environment for each AGN.  The outflows of AGN residing in
  regions of higher density do not fill as large a volume because the IGM
  exerts more pressure on the bubbles than in a uniform, average density
  distribution.  Also, the AGN themselves are not distributed over as large
  a volume of space, as they tend toward higher density regions.
  Therefore, their outflows do not fill as large a fraction of the
  simulation as if they were uniformly distributed.

  \section{RESULTS}
  \label{sec:res}

  We have seen in the previous two sections the general result of our model
  on the filling fraction of AGN outflows.  In this section, we will
  examine the effects of varying simulation box size and resolution, AGN
  lifetime, and the distribution of AGN as determined by AGN bias.  We will
  first conduct convergence studies, to determine the effects of box size
  and resolution on the rest of our analysis.
    
  
  \subsection{Convergence Studies}
  \label{subsection:con}

\begin{figure}[hpt]
    \centering
    \epsscale{.7}
    \plotone{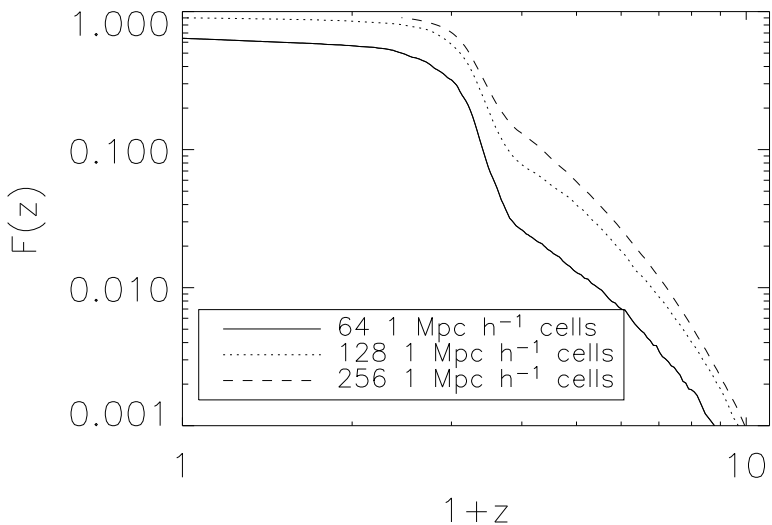}
    \caption{\label{fig:box} Filling fractions for different box volumes.
      Note that due to the computation time required for the $256 \ h^{-1}
      \dim{Mpc}$ length box, we did not calculate the filling fraction all
      the way to $z=0$.}
  \end{figure}

  We ran our analysis on simulations of differing box sizes in order to
  choose the optimal box size for the rest of our studies.  Larger boxes
  are more inclusive, but also much more computationally expensive, and so
  convergence is desirable.  For each box size, we determine the fractional
  volume heated by AGN as a function of redshift using the model for bubble
  growth described in \S \ref{sec:out}.  We have evolved boxes of length
  $64\ h^{-1}\dim{Mpc}$, $128\ h^{-1}\dim{Mpc}$, and $256\ h^{-1}\dim{Mpc}$
  (each with $1 \ h^{-1}\dim{Mpc}$ cells) on a side.  Figure \ref{fig:box}
  shows the filling fraction for each box size.  The filling fraction does
  not vary dramatically between the two larger box sizes (for this reason,
  we did not complete the run for the largest, most expensive box size).
  It appears that we have reached convergence for the $128\
  h^{-1}\dim{Mpc}$ box.  We therefore choose the $128\ h^{-1}\dim{Mpc}$ box
  size for all of our parameter studies.

\begin{figure}[hpt] 
    \centering
    \epsscale{.47}
    \plotone{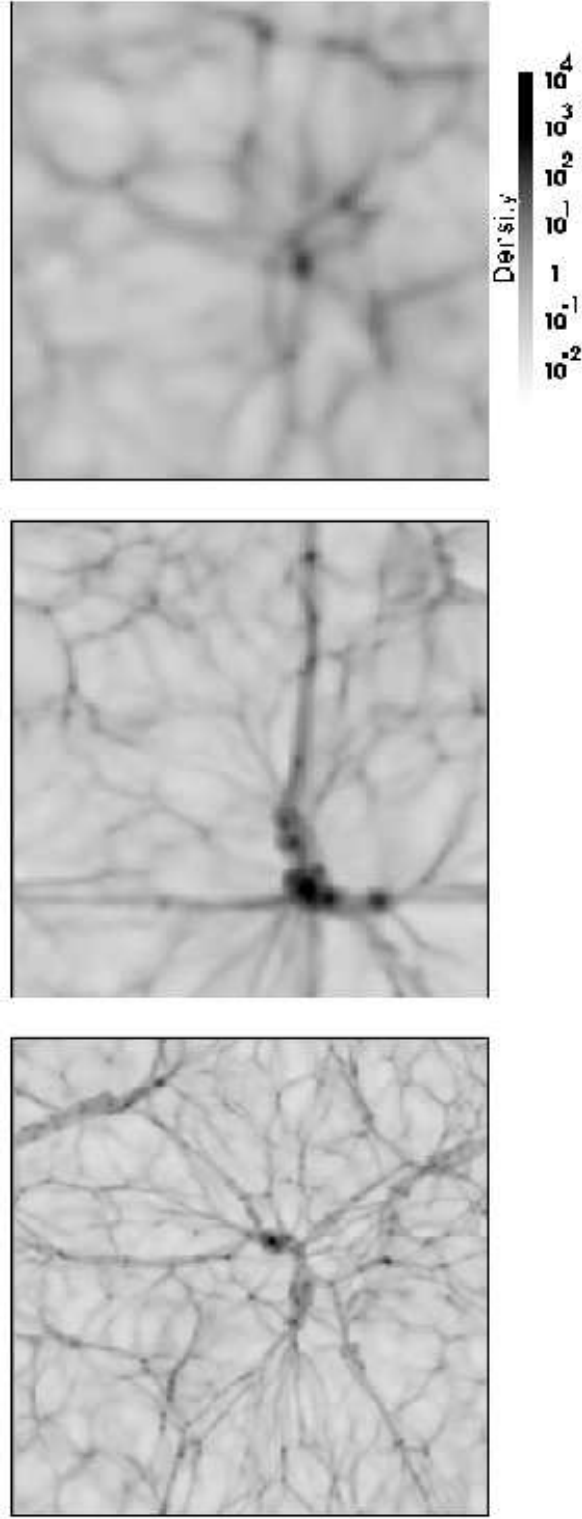}
    \caption{\label{fig:gas} Examples of different resolutions.  The above
    Figures show density distributions at z=0 for different runs with
    differing resolutions.  The upper box has $1\ h^{-1}\dim{Mpc}$ cells,
    the center box has $0.5\ h^{-1}\dim{Mpc}$ cells, and the lower box has
    $0.25 \ h^{-1}\dim{Mpc}$ cells.  All boxes are $64\ h^{-1}\dim{Mpc}$
    across.  A wider range of densities is reached for finer resolved
    boxes.}
  \end{figure}

\begin{figure}[hpt]
    \centering
    \epsscale{.7}
    \plotone{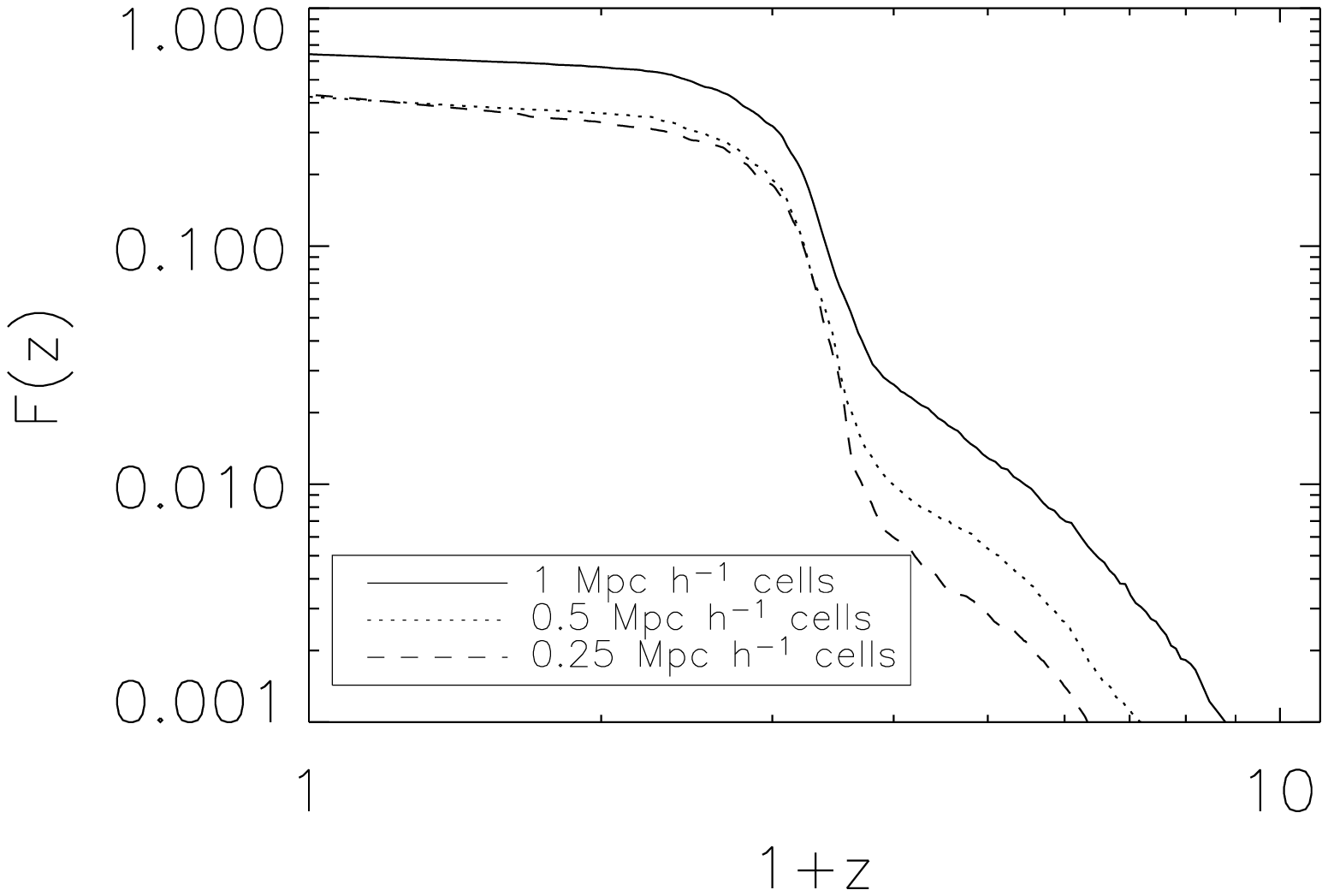}
    \caption{\label{fig:res} Filling fraction for different cell
    resolutions of a box of length $64\ h^{-1}\dim{Mpc}$.  The two most
    finely resolved runs produce very similar results by $z=2$, and so we
    choose $0.5\ h^{-1}\dim{Mpc}$ cells as our fiducial resolution, as it
    is the more computationally inexpensive of the two.}
  \end{figure}

  In addition to box size studies, we also examined the effects of
  simulation resolution on the AGN heated fractional volume.  In
  simulations with finer resolution, individual cells can reach
  significantly higher densities (see Fig. \ref{fig:gas}), which directly
  affects the sizes of the bubbles in our simulations.  The result is a
  different fractional volume affected by AGN depending on resolution.
  High resolution represents a more accurate picture of the simulated
  volume, but like large box size, is more computationally expensive
  because of the larger number of cells.  Therefore, we increase the
  resolution in our simulation box until the filling fraction no longer
  depends on resolution, after which, there is no need for finer
  resolution.  We studied the $64\ h^{-1}\dim{Mpc}$ length box with
  resolutions of $1\ h^{-1}\dim{Mpc}$ cells, $0.5\ h^{-1}\dim{Mpc}$ cells,
  and finally $0.25\ h^{-1}\dim{Mpc}$cells.  We reach convergence in the
  filling fraction very quickly, as shown in Figure \ref{fig:res}.  The
  results for the $0.5\ h^{-1}\dim{Mpc}$ cells and the $0.25\
  h^{-1}\dim{Mpc}$ cells are very similar, and so we choose the faster
  computation, $0.5\ h^{-1}\dim{Mpc}$ cells, for our fiducial resolution.

  \subsection{AGN Lifetime}
  \label{subsection:life}
  
  We have chosen for our fiducial model a constant lifetime for all AGN,
  $\tau_{AGN} = 10^8\dim{yrs}$.  Yu \& Tremaine (2002) examine the
  dependence of lifetime on black hole mass and their results show modest
  variation in lifetime ($\sim 30 - 300\dim{Myr}$) over the range of black
  hole masses of interest here.  They determine AGN lifetime from a
  combination of the luminosity function and black hole number density, and
  their results are comparable to the associated Salpeter times, further
  evidence that black holes accrete most of their mass during their active
  phases.  Our choice of $\tau_{AGN} = 10^8\dim{yrs}$ is consistent with
  their results.  We test the effect of using higher and lower lifetimes as
  well.  We apply a constant lifetime of $10^7\dim{yrs}$ as our lower
  limit.  According to S \& B, $10^7\dim{yrs}$ is an approximate lower
  limit for AGN lifetime, obtained from arguments similar to those of Yu \&
  Tremaine:

  \begin{equation} \label{eq:fon}
    f_{on}(z) = \frac{\tau_{AGN}}{t_{Hubble}} \ge \frac{\Phi(>L_B,
    z)}{n(>M_{BH}, z=0)}\;.
  \end{equation}

  \noindent In the above equation (eq. 22 of S \& B), the AGN lifetime is
  determined from the fraction of galaxies having active nuclei at a given
  redshift ($f_{on}$).  This fraction can be determined by the ratio of the
  comoving number density of AGN (where $\Phi$ is the number density of AGN
  with luminosities greater than $L_B$ at redshift $z$) to the black hole
  number density ($n$).  S \& B assume that the black hole number density
  only increases with time, so that eq. \ref{eq:fon}, with the number
  density evaluated at $z=0$, will give a minimum lifetime.  For our upper
  limit, we use $\tau_{AGN} = 10^9\dim{yrs}$, consistent with Croom et
  al. (2004) who determine an upper limit on AGN lifetime from arguments
  about the growth of dark matter halo mass.  We do not examine redshift or
  mass dependence of $\tau_{AGN}$, because the variation is not significant
  over the range of lifetimes we consider.

\begin{figure}
    \centering
    \epsscale{.7}
    \plotone{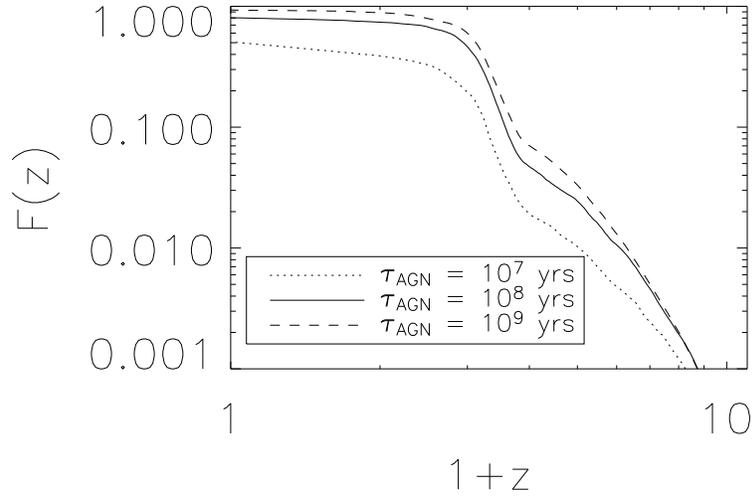}
    \caption{\label{fig:life} Filling fraction for different values of
    constant AGN lifetimes.  Shorter AGN lifetimes imply younger AGN living
    in higher density environments.}
  \end{figure}

\begin{figure}
    \centering
    \epsscale{.7}
    \plotone{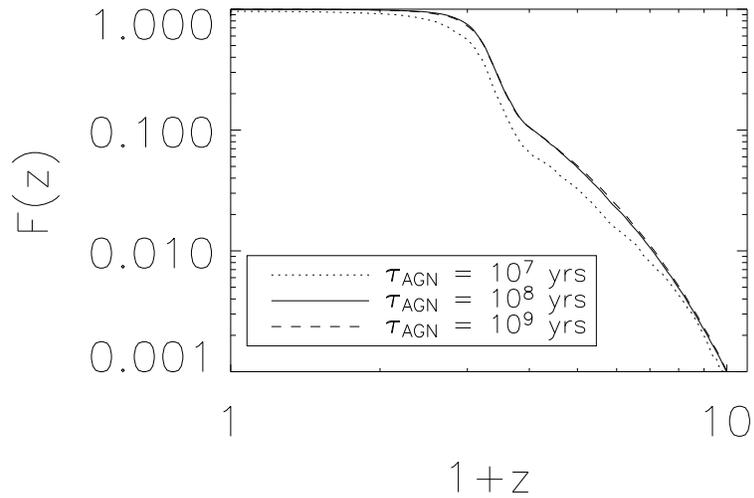}
    \caption{\label{fig:dens1} Filling fraction for different values of constant
    AGN lifetimes for a constant, isotropic density distribution.  Without
    an evolving density distribution, lifetime does not strongly effect the
    filling fraction.}
  \end{figure}

  We find that shorter lifetimes result in a lower filling fraction.  In
  order to remain consistent with the luminosity function in the case of
  shorter lifetimes, many more AGN will become active at later redshifts
  than in the longer lifetime case.  As AGN are born later in the
  simulation, they sit in regions of higher density than AGN born earlier,
  as the density distribution evolves with redshift.  In this analysis, we
  have not included the possibility of recurrent activity associated with
  the same nucleus because the duty cycles of AGN are so low that it would
  introduce a small effect.  Figure \ref{fig:life} shows $F(z)$ for
  lifetimes of $10^7$, $10^8$, and $10^9\dim{yrs}$.  In order to test that
  our interpretation of the effect of evolving density distribution is
  correct, we calculate $F(z)$ for the three values of AGN lifetime under
  the assumption that the density distribution is uniform throughout the
  universe.  Figure \ref{fig:dens1} shows the same range of lifetimes for
  the uniform density case.  As expected, here we do not see the trend with
  lifetime because the density dependence has been removed.  Therefore,
  because of the evolving density distribution, a shorter AGN lifetime will
  result in a lower filling fraction at $z=0$.

  \subsection{AGN Bias}
  \label{subsection:bias}

  Observations show that AGN are biased toward regions of high density,
  with this bias increasing toward higher redshifts.  A simple,
  quantitative implementation of this observed bias is the relation
  $n_{AGN} \propto {\rho_{m}}^{\alpha}$, where $n_{AGN}$ is the number
  density of AGN in units of average number density of AGN, and $\alpha$ is
  the linear bias parameter.  Therefore, $\alpha$ is a measure of the
  correlation between matter density and AGN density.  We examine simple
  constant bias models here as well as one redshift dependent model.  We do
  not, however, examine the possibility of luminosity dependent bias here,
  nor do we distinguish between the bias of radio-loud and radio-quiet QSOs
  despite evidence that radio-loud sources are more strongly clustered.
  Therefore, our bias parameter, $\alpha$, should be considered as an
  average bias of the outflow-producing population of AGN.

\begin{figure}[ht]
    \centering
    \epsscale{.7}
    \plotone{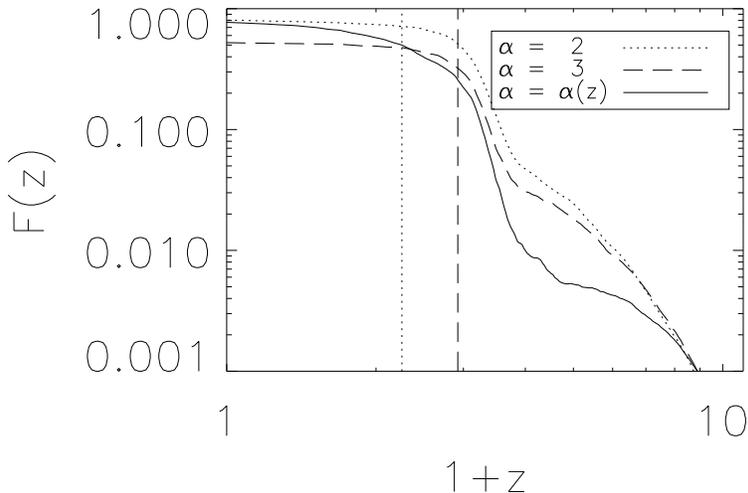}
    \caption{\label{fig:bias} Comparison of filling fractions computed with
      different bias parameters.  The two constant bias parameters are
      shown as well as a z-dependent bias described in \S
      \ref{subsection:bias}.  The two vertical lines are correspond to the
      redshifts at which the functional form of the bias equals each of the
      constant biases ({\it dashed}: $\alpha=3$ and {\it dotted}:
      $\alpha=2$)}
  \end{figure}

  Our fiducial model was run with a constant bias, $\alpha = 2$, consistent
  with the average of many bias models.  Because bias increases with
  redshift, we also ran a model with an increased constant bias of $\alpha
  = 3$.  With a larger bias, AGN will be more tightly clustered to higher
  density regions, preventing their outflows from growing as large, and
  resulting in more overlap between bubbles.  The result is a lower filling
  fraction, as shown in Figure \ref{fig:bias}.

  Since the value of the bias parameter is not constant, but more likely to
  be (at the very least) redshift dependent, we have also tested the
  following redshift dependent model from Croom et al. (2004):

  \begin{equation}
    \alpha(z) = 0.53 + 0.289 (1 + z)^2\;.
  \end{equation}

  \noindent The above is a simple model derived from 2dF data (up to
  $z<2.48$) combined with WMAP and 2dF cosmology.  Rather than extrapolate
  this model over our entire redshift range, for $z \ge 3$ we use a
  constant bias of $\alpha=5.154$, determined by evaluating the above
  equation at $z=3$.  The effects of redshift dependent bias are seen in
  Figure \ref{fig:bias}.  At higher redshifts, AGN are more strongly biased
  toward high density regions than in either of our constant bias models,
  and so the filling fraction is significantly lower.  At lower redshifts
  ($z \lesssim 1$), AGN are even less biased toward regions of high density
  than in our fiducial model, and so $F(z)$ increases approaching $z=0$.

  \section{DISCUSSION AND CONCLUSIONS}
  \label{sec:con}
  
  We have examined the filling fraction of AGN outflows in the context of
  large-scale cosmological simulations, and considered the influence of
  various observationally constrained parameters on the result.  We find
  that the kinetic fraction of outflows need not be very high ($\sim 10\%$)
  for AGN outflows to fill the entire IGM by $z \sim 2$.  Observations of
  gaps in the \lya forest provide possible constraints on the filling
  fraction, that can in turn be used to place constraints on the kinetic
  luminosity.  In our study we have used a luminosity function consistent
  with optical surveys to distribute outflows throughout our simulation,
  however, future studies will have to consider X-ray surveys, which
  predict more faint luminosity AGN.

  Our model employs several simple approximations, but is nonetheless
  instructive.  We have made the assumption that AGN outflows are spherical
  bubbles, propagating into the AGN adiabatically for a short while, until
  reaching pressure equilibrium with their environments.  Spherical
  symmetry is not likely to remain intact once the outflow reaches far
  enough into the IGM.  The outflows will expand away from large-scale
  structures, such as filaments, and into less dense regions.  However, we
  can assume spherical symmetry as an average geometry, and the effects on
  the volume filling fraction are not likely to be very large.  Our
  assumption that radiative cooling can be ignored is justified by the
  timescales involved.  If we were to consider more complex environments in
  our model, such as those of clusters, we would need to include a great
  deal more physics (including cooling) requiring hydrodynamical
  simulations.  While the actual outflow physics and geometry are likely to
  be more complex than we assume, our assumptions provide a good
  first-order approximation of the filling fraction of AGN.

  The parameters we have studied, AGN lifetime, and AGN bias, contribute
  significantly to the AGN filling fraction as expected.  We have assumed a
  particular evolution of the gas density distribution, in which the gas
  density profile is relatively uniform at high-z, and forms high density
  filaments at low-z.  We have tested upper and lower limits for AGN
  lifetime, and find that shorter AGN lifetimes result in a lower filling
  fraction than longer lived AGN due to the evolving gas density
  distribution.  We have examined the effects of different AGN biases on
  filling fraction as well.  Larger bias results in an ultimately lower
  filling fraction.  However, bias likely depends on factors such as
  redshift and luminosity, and the dependence of the evolution of the
  filling factor on bias is likely to be more complicated than in our
  simple scenario.  Changes both in the bias, and in the AGN lifetime
  affect the filling fraction of outflows because of the importance of the
  underlying density distribution and its importance for determining AGN
  distribution and environments.

  We are very grateful to Mitch Begelman for his contributions to this
  paper.  Additionally, we thank Nahum Arav, Jack Gabel, Mateausz
  Ruszkowski, Mike Shull, and John Stocke for helpful discussions.  This
  work was supported by NSF grant AST-0134373 and by the National
  Computational Science Alliance under grant AST-020018N, and utilized the
  SGI Origin 2000 array and IBM P690 array at the National Center for
  Supercomputing Applications.
  

  
\end{document}